\documentclass[]{aa}
\usepackage{graphicx}
\usepackage{subfigure}
\usepackage{amsmath}
\usepackage{amssymb}
\usepackage{txfonts}
\def\kms{\relax \ifmmode {\,\mbox{km\,s}}^{-1}\else \,\mbox{km\,s}$^{-1}$\fi}
\def\ha{\relax \ifmmode {\mbox H}\alpha\else H$\alpha$\fi}
\def\hb{\relax \ifmmode {\mbox H}\beta\else H$\beta$\fi}
\def\hi{\relax \ifmmode {\mbox H\,{\scshape i}}\else H\,{\scshape i}\fi}
\def\hii{\relax \ifmmode {\mbox H\,{\scshape ii}}\else H\,{\scshape ii}\fi}
\def\oiii{\relax \ifmmode {\mbox O\,{\scshape iii}}\else O\,{\scshape iii}\fi}
\def\oii{\relax \ifmmode {\mbox O\,{\scshape ii}}\else O\,{\scshape ii}\fi}
\def\oi{\relax \ifmmode {\mbox O\,{\scshape i}}\else O\,{\scshape i}\fi}
\def\nii{\relax \ifmmode {\mbox N\,{\scshape ii}}\else N\,{\scshape ii}\fi}
\def\sii{\relax \ifmmode {\mbox S\,{\scshape ii}}\else S\,{\scshape ii}\fi}
\def\lha{\relax \ifmmode \mbox {L}_{H\alpha}\else $\mbox{L}_{H\alpha}$\fi}
\def\ldig{\relax \ifmmode {\mbox L}_{DIG}\else ${\mbox L}_{DIG}$\fi}
\def\ls{\relax \ifmmode {\mbox L}_{ Str}\else ${\mbox L}_{ Str}$\fi}
\def\eme{\relax \ifmmode {\,\mbox{pc\,cm}}^{-6}\else \,pc\,cm$^{-6}$\fi}
\def\l{\relax \ifmmode  \lambda\else $\lambda$\fi}

\def\me{$^{-1}$}
\def\arcmin{\hbox{$^\prime$}}
\def\arcsec{\hbox{$^{\prime\prime}$}}
\def\deg{\hbox{$^\circ$}}
\def\fs{\hbox{$^{\rm s}$}}
\def\hms#1h#2m#3s{\relax \ifmmode #1^{\rm h}\,#2^{\rm m}\,#3^{\rm s}
                   \else \hbox{$#1^{\rm h}\,#2^{\rm m}\,#3^{\rm s}$}
                  \fi}
\def\dms#1d#2m#3s{\relax#1\deg\,#2\arcmin\,#3\arcsec}
\def\hmsd#1h#2m#3.#4s{\relax\ifmmode #1^{\rm h}\,#2^{\rm m}\,#3.#4\fs
                      \else \hbox{$#1^{\rm h}\,#2^{\rm m}\,#3#4\fs$}
                      \fi}
\begin{document}

   \title{Propagation of ionizing radiation in \hii\ regions: the effects of 
optically thick density fluctuations}


   \author{C. Giammanco
          \inst{1}
          \and J.~E. Beckman\inst{1,2}
          \and A. Zurita\inst{3}
	  \and M. Rela\~no\inst{1}
           }

   \offprints{C. Giammanco}

   \institute{Instituto de Astrof\'\i sica de Canarias, C. V\'\i a L\'actea s/n,
      38200--La Laguna, Tenerife, Spain\\
              \email{corrado@ll.iac.es, jeb@ll.iac.es, mpastor@ll.iac.es}
         \and Consejo Superior de Investigaciones Cient\'\i ficas, Spain
         \and Dpto. de F\'\i sica Te\'orica y del Cosmos, Facultad de Ciencias, U. de Granada, 
            Avda. Fuentenueva s/n, 18071--Granada, Spain\\
             \email{azurita@ugr.es}
             }

   \date{}

   \abstract{The accepted explanation of the observed dichotomy of two orders of 
magnitude between in situ measurements of electron density in \hii\ regions, derived from
emission line ratios, and average measurements based on integrated emission
measure, is the inhomogeneity of the ionized medium. This is expressed as a
``filling factor", the volume ratio of dense to
tenuous gas, measured with values of order 10$^{-3}$. Implicit in the filling 
factor
model as normally used, is the assumption that the clumps of dense gas are
optically thin to ionizing radiation. Here we explore implications of assuming
the contrary: that the clumps are optically thick. A first consequence is the
presence within \hii\ regions of a major fraction of neutral hydrogen. We
estimate the mean H$^\circ$/H$^+$ ratio for a population of \hii\ regions in the 
spiral
galaxy NGC~1530 to be the order of 10, and support this inference using
dynamical arguments. The optically thick clumpy models allow a significant
fraction of the photons generated by the ionizing stars to escape from their 
\hii\
region. We show, by comparing model predictions with observations, that these
models give an account at least as good as, and probably better than that of
conventional models, of the radial surface brightness distribution and of
selected spectral line diagnostics for physical conditions within \hii\ regions.
These models explain how an \hii\ region can appear, from its line ratios, to be
ionization bounded, yet permit a major fraction of its ionizing photons to
escape. 
}

   \maketitle
%

\section{Introduction:
line emission diagnostics in \hii\ regions}

Emission lines from different ionization states of the most abundant
elements are one of the most valuable and practical tools for 
determining element abundances in galaxies. They are relatively easy 
to detect and quantify. 
The lines emitted by \hii\ regions are intrinsically bright and sharp 
compared with stellar absorption features, so easy enough to pick 
out and measure. If the electron temperature ($T_{\rm e}$) within an \hii\ 
region 
can be determined, it is not too difficult to determine the overall 
abundance of a given element by measuring the relative 
emission line strengths of  lines due to several ions. To estimate $T_{\rm e}$,
the excited [\oiii]$\lambda$4363\AA\  line can be measured and its ratio 
with the $\lambda$4959\AA\ and 5007\AA\ transitions to the ground state, yields a 
good $T_{\rm e}$ value. This works well at intermediate metallicities (e.g. in 
the 
Magellanic clouds) but not at near solar metallicities, where the lower \hii\ 
region 
temperatures lead to a weak lambda 4363\AA\ line, or at very low metallicities 
where 
all the \oiii\ lines are weak. For local \hii\ regions $T_{\rm e}$ can be 
determined well 
from  radio recombination lines (see e.g. Shaver et al. 1983), but for \hii\ 
regions in Milky Way sized galaxies with near solar metallicities either 
detailed 
photo--ionization model fitting, or semi--empirical methods using bright lines 
are used.
One of the most used of these methods involves a calibrated plot of 
[\oii]$\lambda$3727\AA+[\oiii]$\lambda\lambda$4959\AA+5007\AA\ normalized by 
the 
\hb\ emission line strength and plotted versus the oxygen abundance O/H, 
proposed by 
Pagel et al. (1979), and subject to a critical review by V\'\i lchez (1988). 
This plot, and similar plots for N and S, can be used in conjunction with an 
estimate of the intrinsic radiation hardness from the ionizing stars: 
O$^+$,O$^{++}$, S$^{+}$/S$^{++}$ (V\'\i lchez \& Pagel 1988), to measure 
abundances and the abundance gradient 
across the face of a spiral galaxy. An interesting early summary of oxygen 
abundance gradients in 
spirals obtained using these principles can be found in Edmunds (1989).

As early as 1967 Peimbert pointed out that temperature 
fluctuations in \hii\ regions will cause temperature estimates made on the basis 
of 
forbidden line ratios to differ from those made via the ratio of Balmer lines to 
Balmer continuum, and as such will lead to imprecision in any physical parameter 
determination  requiring electron temperatures, notably that of abundances. 
Rubin 
(1969), considered explicitly the effects of temperature fluctuations on 
abundance determinations, showing that for certain excitation conditions, neglect 
of these fluctuations will lead to underestimates of the abundances of elements 
such as oxygen and nitrogen. Rubin (1968), also explored the effects of density 
fluctuations, finding that the smaller and denser are the clumps of dense gas 
within 
a region, the higher the resulting effective electron temperature. He calculated 
models whose volume emissivities in \ha\ and [\oiii]$\lambda$5007\AA\ are 
affected by density and its fluctuation, but these were of the kind represented 
in the 
``filling factor'' approximation which we will consider in some detail below, 
and as such 
are not an especially realistic representation of the density distribution 
within 
an \hii\ region.
In practice, the effect of density fluctuations on the ionization 
equilibrium of a gaseous nebula has been considered essentially only as the cause of 
temperature fluctuations. This was clear in early work on the Orion Nebula by 
Osterbrock \& Flather (1959) and in a subsequent more detailed examination of 
the 
same \hii\ region by Simpson (1973), was  brought out in abundance 
determinations in the Magellanic Clouds by Dufour (1975), and also noticed by 
Krabbe 
\& Copetti (2002) when considering the effect of the high density knots in 
30 Doradus in slightly reducing the local electron temperature in apparent contradiction with Rubin (1968).
In all of these studies, the underlying  assumption is that there is a continuously varying 
amplitude spectrum of fluctuation size and density, although since the emission 
intensity of a 
recombination line scales as the square of the electron density, any 
diagnostic diagram will give the strongest weighting to the densest 
fluctuations. However, relatively little direct attention has been given to the 
effects of the 
density fluctuations themselves on the radiative transfer properties of an \hii\ 
region. Roy et al. (1989) attributed the variation of a number of line ratios 
on small scales within a number of large luminous extragalactic \hii\ regions, to 
small scale density variations within each region, but no direct attention has 
been 
paid to the possible effects of major density fluctuations on the overall 
energy budget of a region. In the present paper we address this problem, using a 
straightforward modelling technique to represent the presence of clumpiness 
within the 
regions.
The clumps in the model show considerably more contrast with their surroundings
than is assumed for a continuous density variation distribution, but the 
model has a density distribution much more like that empirically observed in 
the local neighbourhood (the only place where individual $\sim$1 pc scale clumps 
in 
the ISM can be separately detected and measured). These clumps are also in 
better 
physical accord with the two and three phase models of the ISM which have been 
proposed to explain its observed properties within the Galaxy (Cox \& Smith
1974; McKee \& Ostriker 1977). 

In the following section we give an explanation of the model and its assumptions, 
and in subsequent sections we discuss 
the basic structure of the ionization distribution within the model.
In Sect.~4 we compare
the predictions of the generic radial surface brightness distribution of an 
individual region with the observationally derived distributions for 
\hii\ regions in external galaxies. Comparison of some predicted emission line 
ratios 
for this and previous models with observed line ratios, are discussed in 
Sect.~5, and finally, in Sect.~6 we analyze the 
important 
implications of this model for the observationally derived masses of gas within 
\hii\ 
regions. In the conclusions we bring out in what ways these clumpy models are 
qualitatively different from both the classical Str\"omgren model with 
homogeneous gas, 
and its derivatives obtained using the standard ``filling factor'' approach. 
\section{Inhomogeneities: the classical filling factor or optically thick 
clumps?}
 The possible effects of density inhomogeneities on the propagation of ionizing radiation in \hii\ regions were
first discussed in some detail by Str\"omgren (1948). However
 the standard way to take into account the fact that the ISM is highly 
non--homogeneous on scales of parsecs is to scale the optical paths and the 
volume emissivity by a filling factor, following the initial 
prescription of Osterbrock \& Flather (1959), which we will term here the FF 
approximation.
The technique assigns a fraction of the total volume of an \hii\ region 
to relatively dense gas, with the remaining volume taken to be of 
negligible density.
The fraction occupied by dense gas is termed the filling factor, and it 
takes values of order 10$^{-3}$ in typical observed cases. The need to treat the 
volume of a region in this way is because the measurement of the in situ 
electron 
density using line ratios either at radio wavelengths (e.g. Dravskikh \&
Dravskikh  1990) or optical wavelengths (e.g. Copetti et al. 2000) gives values 
of order a few times 10$^{2}$ cm$^{-3}$, whereas the mean electron 
density obtained from the emission measure in \ha\
is of order 1--2  cm$^{-3}$ (Rozas et al. 1996). 
In the form introduced by Osterbrock \& Flather (1959) and in general use for 
dealing with the transfer of Lyman photons in \hii\ regions, the underlying 
assumption is that the dense clumps are small enough to be optically thin, so 
that 
each clump is fully ionized. The overall effect is a statistical modification 
of the classical Str\"omgren sphere structure, in which radiation from 
the central ionizing stars ionizes material within a given radial distance 
from the stellar cluster and is effectively fully absorbed within the \hii\ 
region, except for the minority of cases where either the placental cloud is 
sufficiently small for some radiation to escape, or the cluster is 
formed eccentrically within the cloud, and thus near the cloud edge in one 
direction.
Although the physical conditions for recombination line formation are 
biased towards the high densities, since these emission line strengths are 
proportional to the square of the electron density, and hence this model in 
which 
only the dense clumps are considered, gives many valid results, we believe that 
a better description of the transfer is given by using a medium with 
clumps whose optical depth is high. Rather interesting evidence for this 
is found within the local ISM where a detailed 
examination of structure is possible within a zone which lies within a major 
\hii\ 
region based on the OB associations towards Scorpio and Centaurus. Here Trapero 
et al. (1992, 1993) found that more than half the mass of the ISM within 300 
pc of the Sun is in the form of dense ($\sim$ 100 cm$^{-3}$)  compact clouds 
with 
characteristic sizes of order a few parsecs, while the remainder forms a warmer 
substrate 
with densities lower than 1 cm$^{-3}$. Inspection of optical images of \hii\ 
regions within the Galaxy, e.g. the Orion Nebula, confirms this picture of 
ionized 
material in dense blobs or clumps several parsecs across, separated by larger 
volumes of 
less dense gas. In fact, this structure conforms to the rather clean 
separation into cool and warm phases first proposed by Field, Goldsmith \& 
Habing 
(1969) and later extended to three phase models which include a hot component 
by Cox \& Smith (1974) and by McKee \& Ostriker (1977). A particularly clear 
theoretical underpinning of these multi phase models is given in the 
classical text by Spitzer (1978), who explains in terms of detailed balance, why 
certain ranges of density and temperature in the ISM are unstable, and how this 
leads to a geometrical differentiation of the stable phases. The transfer 
properties of an \hii\ region with a clumpy structure in which there is a clear 
phase 
separation, with correspondingly large well defined clumps are significantly 
different from  those of a traditionally modelled clumpy region where the 
structure is 
not scaled, and it is this difference which we explore here. We will term 
the traditional model the FF model, and our model the ``clumpy'' model, since 
the density  and volume contrasts in our model are greater. The clumps in our 
model, 
and in the \hii\ regions themselves, are optically thick to the Lyman continuum 
(Lyc) and this has significant effects, e.g. implies that the Str\"omgren sphere 
approximation loses validity, and that traditional diagnostics for photon escape 
such 
as those applied in  McCall, Rybski \& Shields (1985) must be 
re-quantified.

To give an analytical first order approach which is easy to follow we 
first take our clumps to be uniform spheres, with a characteristic size $d$ and 
a cross--section for radiation interception $\sigma$. A clump situated at a 
distance $R$ from the centre of a (spherically symmetric) region intercepts a 
fraction $\sigma /4\pi R^2$ of the ionizing radiation, and for a number density 
of 
clumps $n$, the clumps in a shell of thickness $\delta$ between $R$ and 
$R+\delta$ from 
the centre, intercept a fraction of the radiation $4\pi R^2\delta n (\sigma 
/4\pi R^2)$
i.e. $\delta n \sigma$. We term this fraction $f$, which represents the 
probability that a photon is absorbed within this shell, i.e. the fraction of 
photons absorbed within the shell. With the simplifying assumption that the 
number density of clumps is not so large that there is a significant 
probability that a ray from the centre of the region cuts more than one clump on 
its 
path to the surface, the probability $e$ for a photon to escape from the region 
is given by
\begin{equation}
 e = (1-f)^N		 
\label{e}	
\end{equation}
where $N$ is the total number of shells. We are particularly interested in 
the production of \ha\ by recombination in a process which can be 
considered as a down-conversion of Lyc photons, terming the intensity of \ha\
emission by the $i$'th shell from the centre as $I_i$. We can formulate this in 
terms 
of the intensity, $I^{fill}_i$, which would be produced within a filled shell 
which 
has a thickness $d$, multiplied by the fraction of the shell which 
absorbs photons, i.e. the fractional area occupied by dense clumps, 
giving
\begin{equation}
I_i = I^{fill}_i \, f (1-f)^{(i-1)}
\label{Ii}	
\end{equation}
		 
Both equations, \ref{e} and \ref{Ii}, use a thin spherical shell approximation, 
which will not be valid near the centre of the region, but in any case, if $n$ is 
small the contribution of that shell to the total emission can be neglected, 
not only for simple geometrical reasons, but because in the innermost 
few parsecs the clumps will be dissipated by a combination of stellar winds and 
radiation.
For simplicity we will set the shell thickness as the size of a single 
clump, and we can thus neglect the innermost shell in any calculations. The 
fractional error in doing this for a shell where the clumps were not dissipated 
would be just $f$, (and we will see that this ranges from 0.001 to 0.01 so this 
would be no problem) but in practice the error will be smaller than $f$. Even if 
our computation were incorrect for the first three or four shells, the 
error implied would be small. In particular, if the clumps in the second shell 
were 
either dissipated or fully ionized, our error would again be of order $f$, and 
this problem will not be important, even in regions with the most luminous 
ionizing stars, for shells beyond the third or at most the fourth, for clumps of 
scale size $\sim 1$ pc.

It is instructive to use the spherical clump approximation to derive
some quantitative relations associated with the filling factor, i.e. 
the fractional volume occupied by dense gas. For spherical clumps of radius 
$r$, for which $\sigma$ is $\pi r^2$ we find
\begin{equation}
f = \pi r^2 n				
\end{equation}
and setting the shell thickness delta to be the clump diameter ($\delta=2r$)
we obtain
\begin{equation}
f = (3/2) v_c  n						
\end{equation}
where $v_c$ is the volume of a clump. Let us define a geometrical filling 
factor $\phi_G$ as the fraction of the total volume of the \hii\ region occupied 
by the clumps and so we find 
\begin{equation}
\phi_G = (4/3) r f/\delta
\label{convertion}			
\end{equation}
and if we put $\delta = 2r$ we have 
\begin{equation}
f = (3/2) \phi_G
\label{convertion2}						
\end{equation}

It is important to be explicit here about the method normally used to 
infer filling factors of dense gas in \hii\ regions. Following Osterbrock (1989, 
p. 153), 
the luminosity L emitted by an \hii\ region whose dense clumps have a 
total volume $V_c$ and electron density $N_{\rm e}$, in a given spectral line is
\begin{equation}
L = k {N_{\rm e}}^2 V_c
\label{def_fillingfactor}					
\end{equation}
where $k$ is a constant which depends only on which emission line is 
being observed, and not (to first order) on the physical conditions. If 
we assume that we are dealing with an \hii\ region whose total volume is $V$, 
then the ratio $V_c/V$ is the filling factor, $\phi_\circ$ so we can express 
Eq.~\ref{def_fillingfactor}  as
\begin{equation}
L = k {N_{\rm e}}^2 \phi_\circ V  
\end{equation}
which for the case of \ha\ enables us in principle to determine 
the filling factor $\phi_\circ$ using
\begin{equation}
\phi_\circ = k^\prime L_{H\alpha} / ((4\pi/3) R^3 {N_{\rm e}}^2)			
\label{phi_o}	
\end{equation}
where $R$ is the radius of the \hii\ region, and the constant $k^\prime$ takes 
the 
value  appropriate for the \ha\ emission line. We note especially that in an 
FF model the two filling factors, $\phi_G$ and $\phi_\circ$ are equal, since 
all clumps are fully ionized, but in the present clumpy model this is not the 
case. We will see that in general  $\phi_\circ <  \phi_G$ since most clumps are 
not 
fully ionized, and the difference between these quantities can be easily 
an order of magnitude. In addition, deriving $\phi_\circ$ using an integrated 
emission 
line luminosity and Eq.~\ref{phi_o} implies obtaining a reliable value for the 
radius of the \hii\ region, which is not the most accurate of observations 
because the projected surface brightness of a region blends into the diffuse 
background, so that any measurement which relies on establishing a clear 
boundary 
entails errors. We will discuss these points further, later in the paper.

\section{Calculations using FF and clumpy models}
\label{FF_C}
In order to illustrate both qualitatively and quantitatively the 
differences which arise using the two types of models we have generated 
specific models to compute line strengths and ratios under defined 
conditions.
All these models rely on the CLOUDY suite of programs (see Ferland et al. 1998). 
For 
the FF models we have used CLOUDY in a standard way. We have generated a 
tree family of models of different metallicity with a set of stars: 3, 10, and 
30 O3 stars at the 
centre, a mean gas density of 100 cm$^{-3}$ and filling factors of 10$^{-3}$, 
10$^{-4}$,  
values designed to match the optical filling factors obtained observationally 
(e.g. Rozas et al. 1996). The outer radius of the region is 
a parameter which can be varied, so that if we start with low values and 
progress to high values we go from a density bounded to an ionization bounded 
regime.
We used essentially solar metallicities, as specified in the manual of 
CLOUDY, and for the purpose of streamlining the calculation we simulated an O3 
star by a black body with a temperature of  51230 K, and  
an ionizing flux 10$^{39.31}$erg~s\me\ (Vacca et al. 1996).

\begin{figure}
\resizebox{\hsize}{!}
{\includegraphics{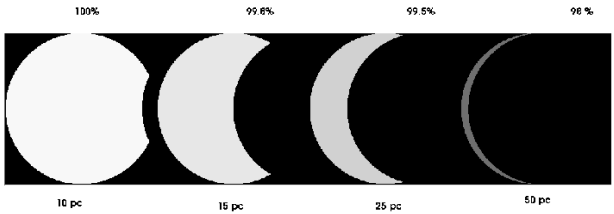}}
\caption{Equatorial section of a spherical clump with density 100 cm$^{-3}$ and 
radius 1 pc
illuminated by an ionizing source and placed at different distances from it.
The light area represents the ionized volume  of the clump (seen in
cross--section) and the dark area represents the volume which remains neutral. 
Gray areas show partial ionization.
In this example, obtained with CLOUDY, we used an ionizing source  emitting
10$^{42}$erg~s$^{-1}$ yielding conservatively high clump ionization.}
\label{grumo_distancia}
\end{figure}

Since the clumpy models are based on clumps of denser gas whose 
properties: size and gas number density, are defined as inputs, we first 
made a set of simple tests in which a clump of radius 1 pc and density 
100~cm$^{-3}$ was illuminated by an ionizing source at different distances. 
The source was given an ionizing luminosity of 10$^{52.5}$~photons~s\me\ 
which corresponds to almost 300 O3. This is a very luminous source, so that
our conclusions about the ionized fractions of the clumps at varying distances from the source
are conservative.
We can see the result of this test in 
Fig.~\ref{grumo_distancia}, where the fractional volume ionized, seen in 
cross--section, 
and the degree of ionization, are shown as functions of the distance 
of the clump from the source.

It is obvious at a glance that at distances further from the source 
than some 10~pc, a clump is essentially opaque to the ionizing radiation, 
and in terms of simple transfer can be virtually considered a black disc. 
However with the filling factors found observationally, the net effect of a 
clumpy 
region where the gas is concentrated into these clumps will be to act as a 
``radiation sieve'' allowing a major fraction of the ionizing photons to pass 
between the clumps and escape. It is also clear that at distances beyond some 
20~pc 
from the source the major fraction of a clump is shielded and left 
unionized.
Depending on the details of the distribution of the clumps within the 
\hii\ region, the  net effect of integrating unionized and ionized 
clump masses will be that a relatively small fraction of all the 
clump mass in the \hii\  region will be ionized. It is this fraction 
which is detected as the ``optical'' filling factor. A useful canonical 
estimate of the ratio of neutral to ionized gas 
in an \hii\ region of radius 150~pc, i.e. one of the ultraluminous 
``giant'' \hii\ regions, is a factor 10, as we will show empirically in section 6
    below, so that in our clumpy models we use a 
``geometrical'' filling factor of 10$^{-2}$, 10$^{-3}$ (including both H$^\circ$ and H$^+$) which in broad terms 
corresponds to an optical 
filling factor of 10$^{-3}$,10$^{-4}$ (including H$^+$ only).

 In the computations made 
using the clumpy models we have used the formalism implied in Eq.~\ref{Ii} in 
conjunction with the CLOUDY program of continuum and line transfer (Ferland et 
al. 1998).

\section{Differences in the radial distributions of ionizing
and recombination photons between the FF and clumpy models}

\begin{figure}
\resizebox{\hsize}{!}
{\includegraphics{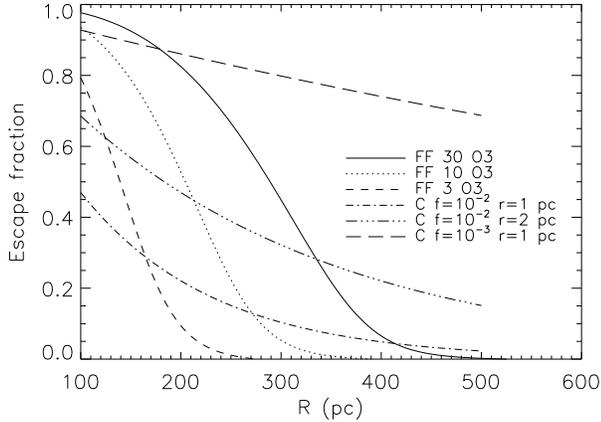}}
\caption{Dependence of photon escape fraction on the 
\hii\ region radius for different models. In classical FF models, 
the escape fraction depends on the radius, on the filling factor, and on the 
strength of the ionizing radiation 
(we have considered ionizing clusters with 3, 10, and 30 O3 type stars, and a 
filling factor of 10$^{-4}$). In contrast, in the clumpy models the dependence is 
only geometrical; in other 
words, the escape fraction depends on the radius of the region, and on  the 
radius and density of the clumps. We show 
results of three clumpy models (labeled with C) assuming  different  filling 
factors 10$^{-2}$, 10$^{-3}$ and  
clump radii (1 and 2~pc).
}
\label{escape_fig}
\end{figure}

A key qualitative difference between FF models and clumpy models is 
the prediction of the fraction of ionizing photons which escape from an 
\hii\ region. In an FF model, as in a classical Str\"omgren uniform model, 
there is a clear distinction between density bounding and ionization bounding, 
which depends only on the mean gas density and the modelled radius of the 
region. In 
slightly simplified terms once a region reaches a specified size for its density 
distribution it will be ionization bounded, so no photons escape, while 
for smaller radii it will be density bounded. In a clumpy model this clear 
distinction is not maintained, since a fraction of ionizing photons 
escapes from essentially all clumpy models with reasonable physical parameters, 
and the only question is 
to quantify that fraction (Eq.~\ref{e}). In other words, all 
regions modelled using a clumpy model will be, de facto, density bounded. 
However, in each individual clump (except for the minority of clumps close to the 
ionizing source) all photons incident on the clump are absorbed. Thus the 
emission line ratios for lines produced by down-conversion within the clump will 
be 
those for ionization bounded regions, and as essentially all the line 
emission from within a region comes from the clumps, any diagnostic diagram 
using line 
ratios will give an ionization bounded signature. In other words clumpy \hii\ 
regions will always be diagnosed as ionization bounded, although in terms of 
photon escape they behave as density bounded.
An illustration of these different types of behaviour is given in 
Fig.~\ref{escape_fig} where we have plotted escape fraction against \hii\ region 
radius for three FF models and three clumpy models.
The difference is clear. The FF models show
behaviour akin to that of a classical Str\"omgren sphere. The fraction of 
photons escaping falls quite sharply as the radius of the \hii\ region reaches a 
critical value, which depends on the luminosity of the ionizing stars and the 
mean density of the gas (in the models illustrated here this latter 
parameter was maintained constant so the effect is not seen). The clumpy models, 
labeled with C, show 
quite different behaviour, which is essentially geometrically dominated. The 
escape fraction for an \hii\ region for a given source luminosity falls off 
steadily as the radius of the region rises, but there is no cut--off radius, as 
the fall--off is determined basically by the geometrical cover factor of the 
clumps, each of which is optically thick. In fact the plots for these 
models slightly underestimate the escape factor, as they have assumed no 
overlapping between clump cross sections for photon blocking.

\begin{figure}
\resizebox{\hsize}{!}
{\includegraphics{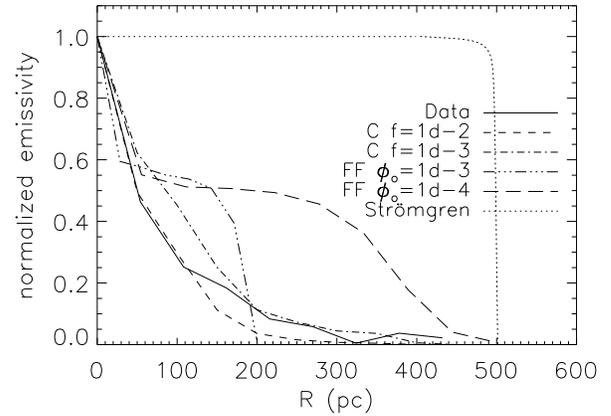}}
\caption{ Observationally derived \ha\  radial emissivity profile of a bright \hii\ region of 
the barred spiral galaxy NGC~1530 (solid line).
The profile has been normalized to facilitate its comparison with the profiles 
predicted by clumpy and FF models.
The spatial resolution  of of the four theoretical models has been degraded to 
match the resolution of the \ha\ image of 
NGC~1530 (approx. 50~pc per pixel). We can see that the clumpy models produce 
concave profiles, similar to the observations. On the
contrary, FF models, give rise to convex profiles. The dotted line represents an 
ideal Str\"omgren model, whose 
ionization fraction is constant and equal to 1 inside the \hii\ region volume, except close to the boundary
of the Str\"omgren sphere.
}
\label{ajuste}
\end{figure}

As a direct observational test of the differences in the predictions 
of FF and clumpy models we have looked at radial profiles in surface 
brightness of \hii\ regions (Rozas et al. 1998), a test which makes best use of 
the 
spherical symmetry incorporated into our simplest clumpy models. 
We selected one of the largest isolated \hii\ regions with 
large radii in the galaxy NGC~1530 for which we had continuum 
subtracted \ha\ images with good signal to noise ratio, measured 
its radial surface brightness profile in  \ha, and from this
profile derived the distribution of volume emissivity in \ha\ 
as a function of radius within the region. The radial plot of this 
contribution function is shown in Fig.~\ref{ajuste}. In the same figure we 
give the predictions of four models, for which the luminosity of the 
central source was scaled to yield the observed
\ha\ luminosity, to a first approximation: two FF models with different 
assumed filling factors and two clumpy models also with different geometrical 
filling 
factors. It is clear from the figure that the FF models give contribution 
functions 
which do not fit those of the observed regions. Although the radial profiles of 
these models differ depending on the filling factor, they have in common a 
convex 
form, with a rather sharp fall off towards the edge of the region. The clumpy 
models, on the other hand, show convex profiles, and give far better agreement 
with those derived from the observations. The \ha\ volume emissivity for a uniform Str\"omgren sphere is
included for comparison.	

\section{Predictions of line strengths and ratios: diagnostic diagrams for 
density bounding}
\newcounter{figmas}
\addtocounter{figmas}{4}
\renewcommand{\thefigure}{\arabic{figmas}}
%
%
\begin{figure*}
\label{mccall}
\resizebox{0.48\hsize}{!}{\includegraphics{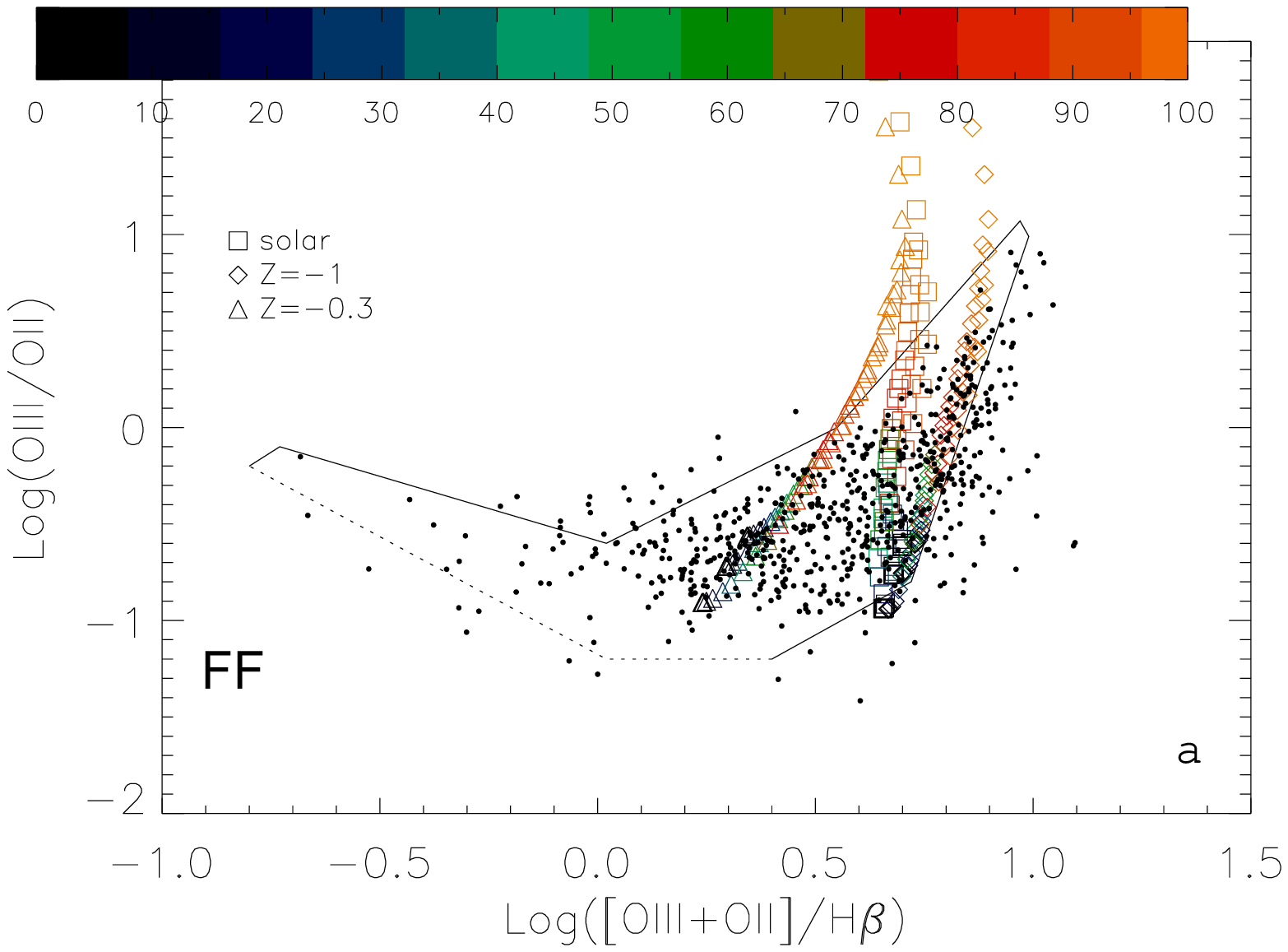}}
\resizebox{0.48\hsize}{!}{\includegraphics{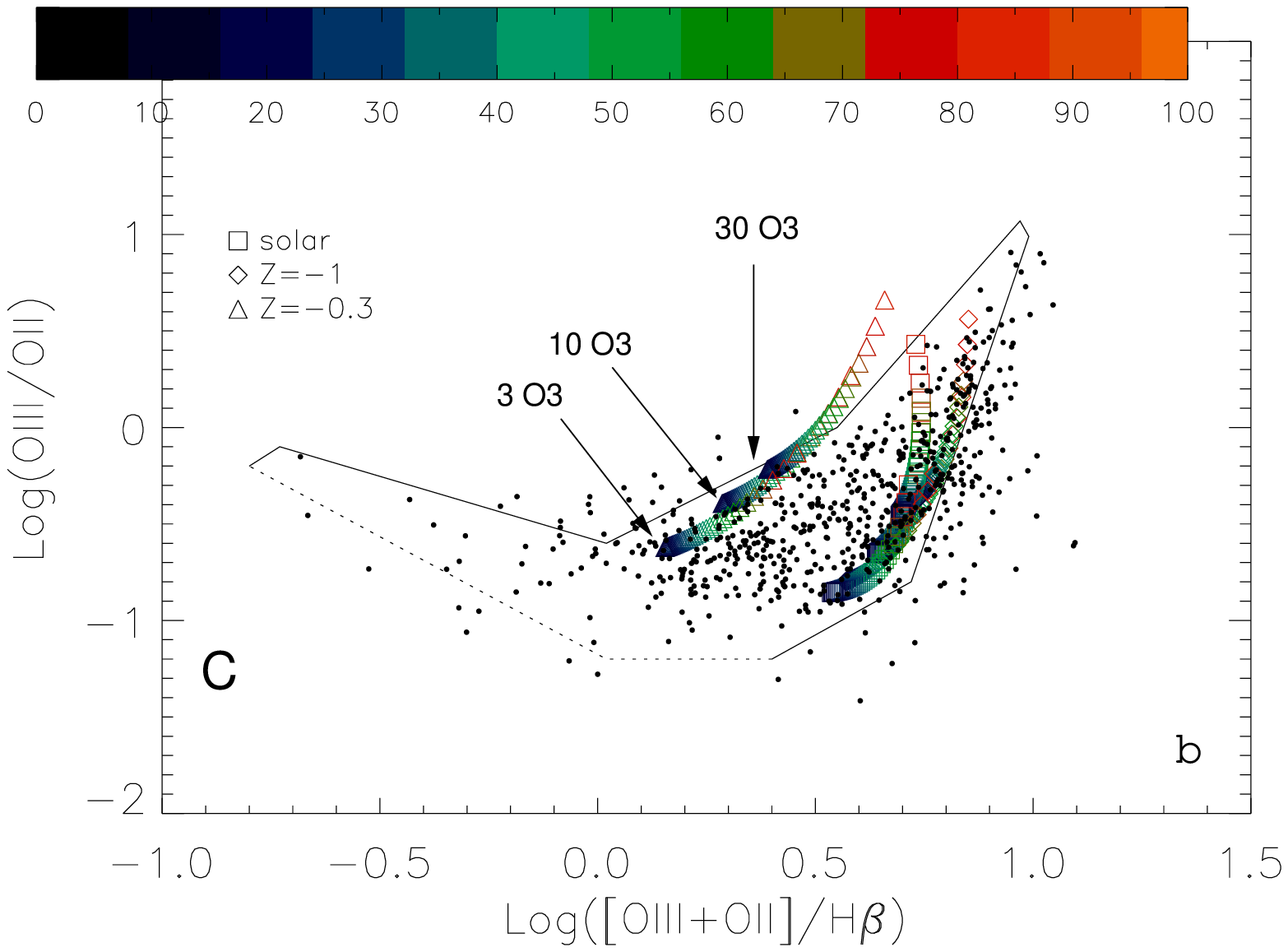}}\\
\vspace{-0.5cm}
\caption{Density bounding diagnostic diagram (McCall et al. 1985). Solid lines 
delimit the area occupied by McCall's data
and the dotted line indicates where this limit is least certain. Data points 
correspond to \hii\ region observations, which have been obtained 
from Cedr\'es \& Cepa (2002), Van Zee et al. (1998),  
Kennicutt \& Garnett (1996), 
Walsh \& Roy (1997) and Roy \& Walsh (1997). Colour symbols refer to models (FF 
in {\bf a.} and clumpy in {\bf b.})
assuming  3, 10  and 30 ionizing  O3  type stars and  3 different abundances: 
 0.1 solar, 0.5 solar, and solar. Colours indicate the escape fraction  of 
ionizing photons in each case. We can see that
all models, even those with a strong escape fraction, are located in the areas 
covered by the observational data.
}
\label{mccall}
%
%
\vspace{0.5cm}
\addtocounter{figmas}{1}
\resizebox{0.48\hsize}{!}{\includegraphics{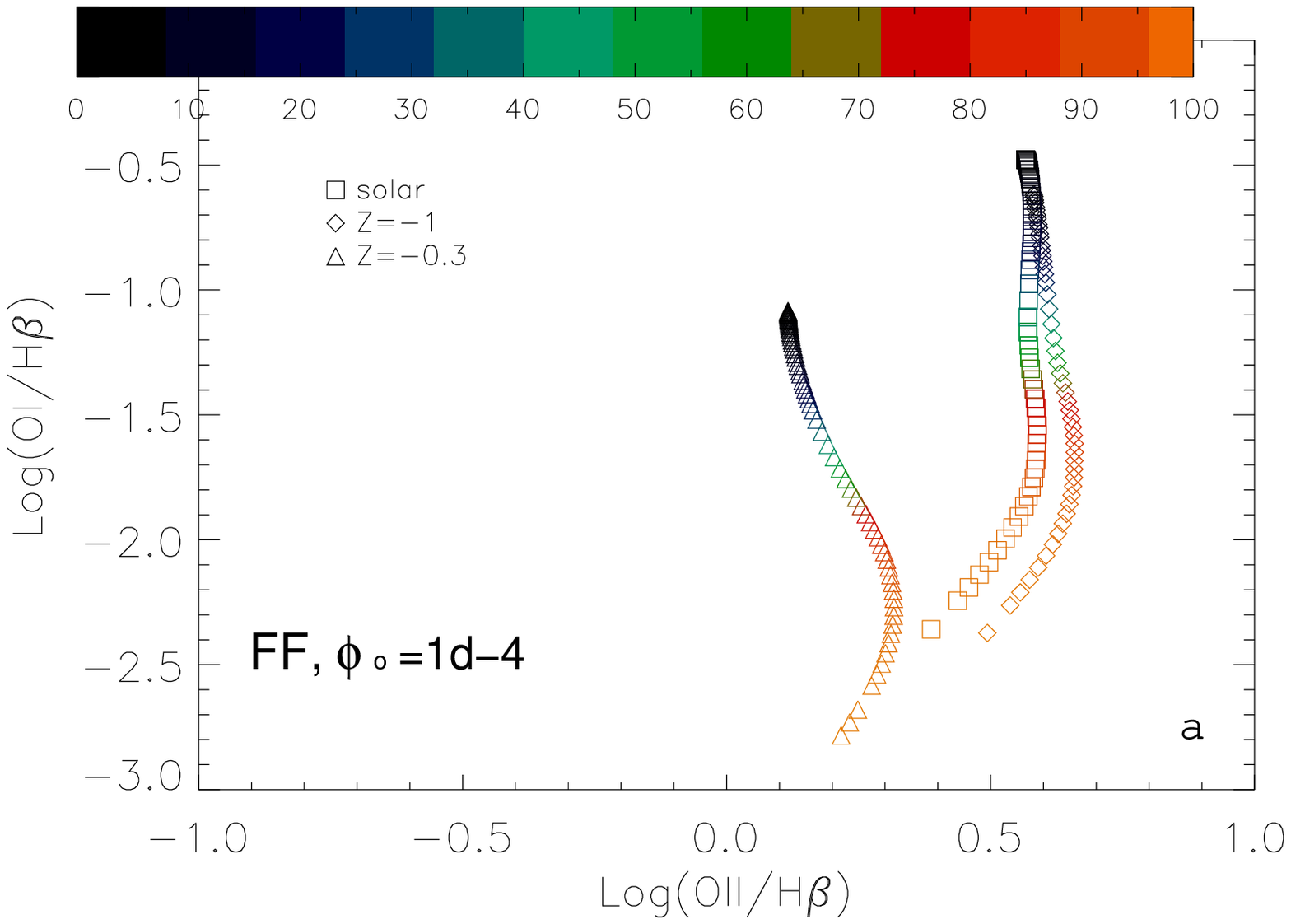}}
\resizebox{0.48\hsize}{!}{\includegraphics{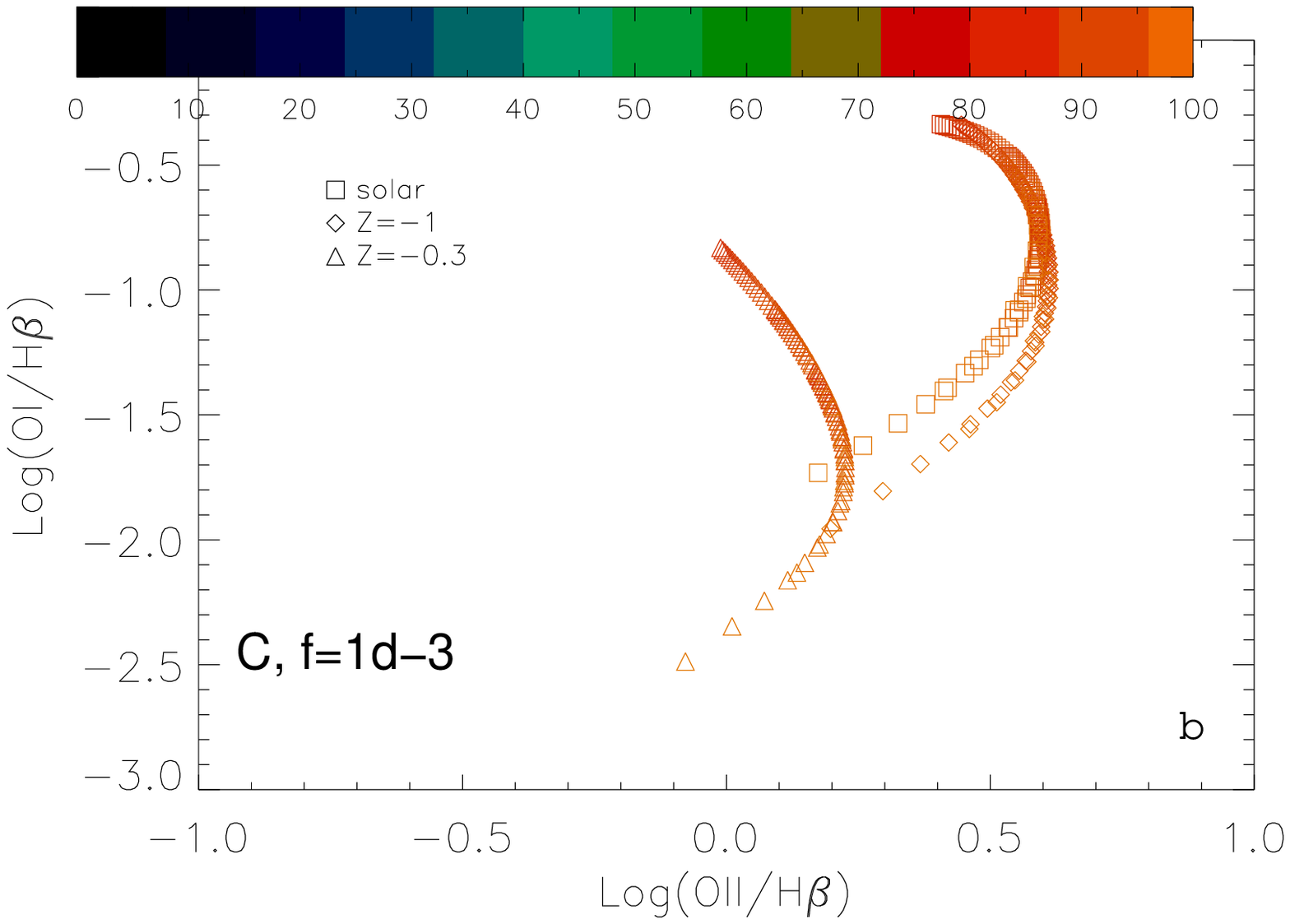}}\\
\resizebox{0.48\hsize}{!}{\includegraphics{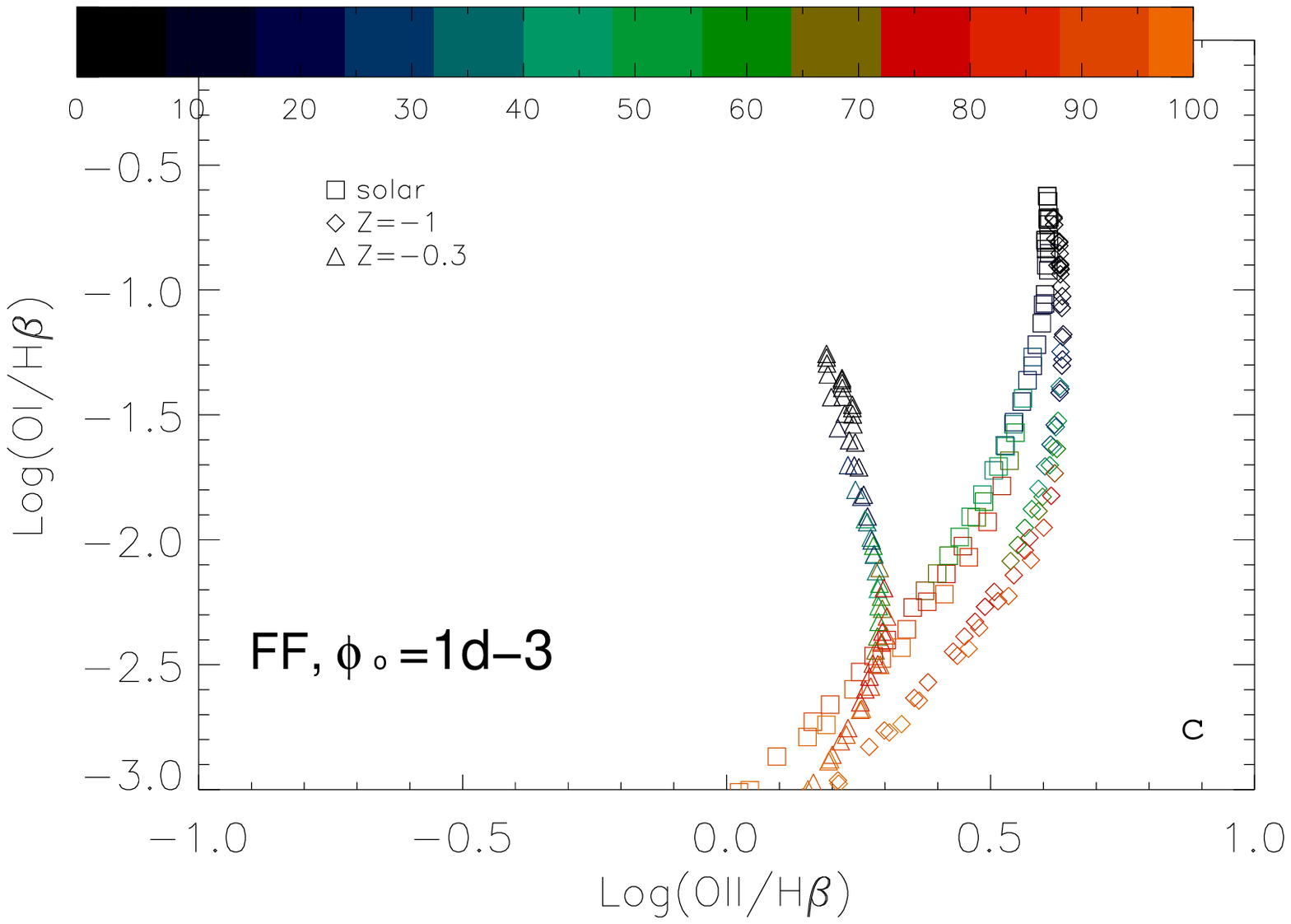}}
\resizebox{0.48\hsize}{!}{\includegraphics{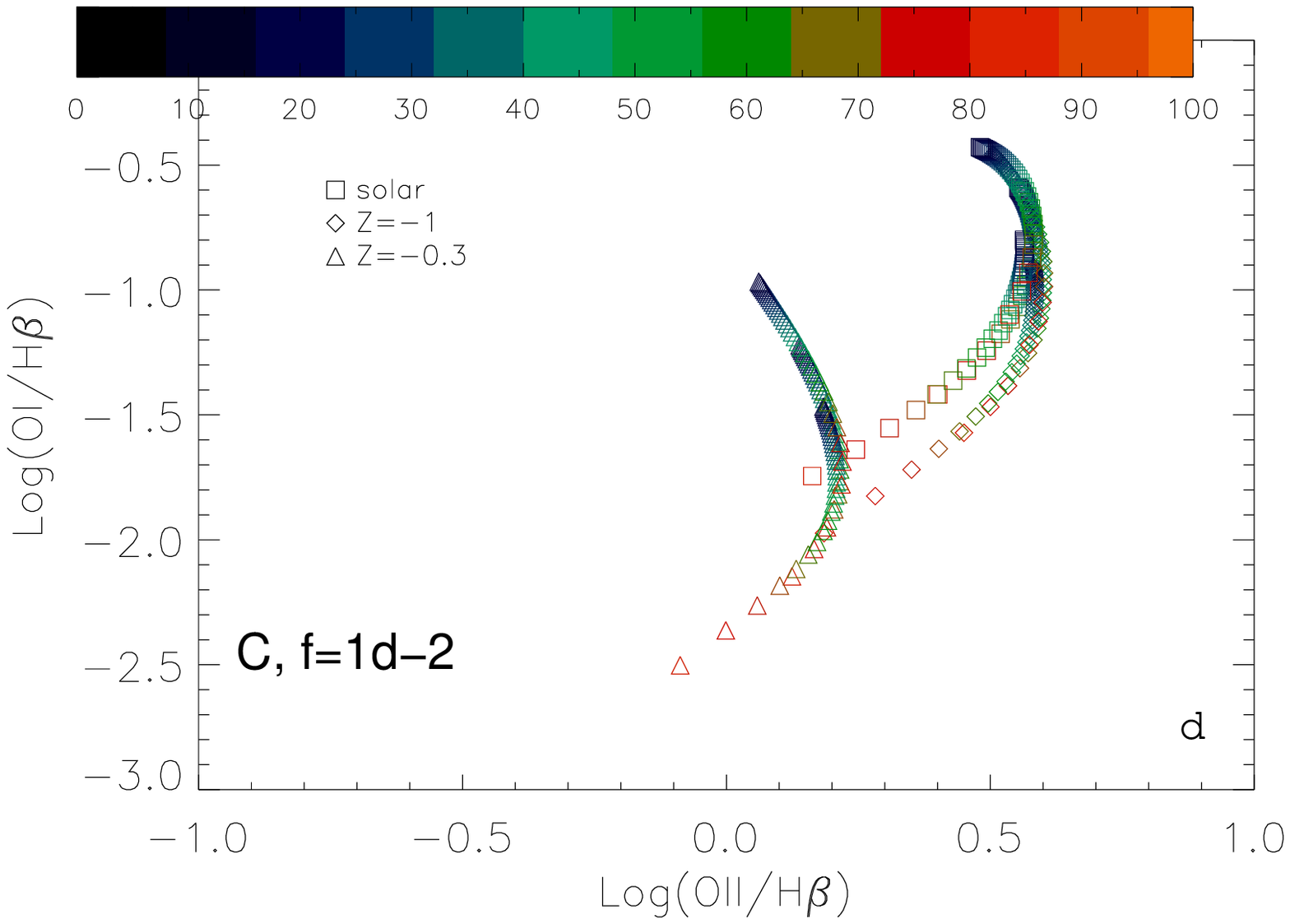}}\\
\vspace{-0.5cm}
\caption{Density bounding diagnostic diagram: [\oi]$\lambda$6300\AA\ vs. 
[\oii]$\lambda$3727\AA\ (cf. Iglesias-P\'aramo \& Mu\~noz--Tu\~n\'on 2002). 
{\bf a.)} Sequence of FF models 
obtained for an optical filling factor 
equal to 10$^{-4}$.  {\bf b.)} Clumpy models for $f=10^{-3}$.
 {\bf c.)} Same as 
a. but for  $\phi_\circ=10^{-3}$. 
{\bf d.)} Same as b. but for  $f=10^{-2}$. The plots show that as the radius 
increases, the data points move 
towards the top-right region of the plots (the ``ionization--bounded'' area) and 
the escape fraction decreases. For clumpy models,
\hii\ regions with high escape fraction (see Fig~\ref{casiana}b for $f= 10^{-3}$)
occupy the same general parameter space as
models with moderate escape fraction (Fig~\ref{casiana}b for for $f= 10^{-2}$). 
For the convertion of $f$ to the geometrical filling factor $\phi_{G}$ see Eqs.~\ref{convertion} and~\ref{convertion2}
($f$ is of order $\phi_{G}$).}
\label{casiana}
\end{figure*}
A classical test for density bounding in \hii\ regions is the plot  
of [\oiii]/[\oii] vs. ([\oiii]+[\oii])/\hb, 
proposed by McCall et al. (1985). Examination of this plot for a set of 
extragalactic \hii\ regions,
led them conclude that most \hii\ regions are ionization bounded. Other authors 
(Roy \& Walsh 1997; Kennicutt et al. 2000) have reached the
same conclusion based on similar analysis.
However, there is good 
evidence (Ferguson et al. 1996; Zurita et al. 2000; Rela\~no et al. 2002) that a major 
fraction, of order half, of the ionizing photons are escaping 
from the \hii\ regions of external galaxies, and ionizing their
diffuse interstellar media. This evidence is based on two types of 
measurements.
Firstly the integrated diffuse \ha\ measured over the full face of a 
galaxy is found to depend linearly on the integrated \ha\ luminosity of the 
most luminous \hii\ regions (above a specified lower limit in \ha\ luminosity) 
in 
that galaxy (Beckman \& Zurita 2004), suggesting very strongly that escaping 
Lyman continuum photons from the \hii\ regions are responsible for ionizing the 
diffuse 
medium, and by inference are escaping from  the \hii\ regions themselves. 
Secondly a careful model of a specific galaxy (NGC~157) in which all 
the \hii\ regions, with measured \ha\ luminosities and positions 
were taken as sources of ionizing photons, which were 
allowed to propagate across the galaxy, gave excellent agreement between 
the modelled geometrical distribution of the diffuse \ha\ emission from the 
whole galaxy and the observed distribution (Zurita et al. 2002). The fraction of 
photons which need to escape is at least 30\% of those emitted by the OB stars 
within 
the regions. 
We have predicted some of the line ratios previously used for density bounding 
diagnostics 
using our clumpy models and compared these predictions with those of the 
classical FF models.
In Fig.~\ref{mccall} we show the large data set presented by several authors 
(including McCall et al. 1985)
in the $\log$([\oiii]/[\oii])--$\log$(([\oiii]+[\oii])/\hb) parameter space,
together with the predictions of the FF (\ref{mccall}a) and clumpy (\ref{mccall}b) models 
for different metallicities. We can see that in both cases, the observations 
can be reproduced by  models with strong ionizing photon leakage (indicated 
by color scale) 
and models with negligible escape of ionizing photons.
We can also infer two types of conclusions from this comparison.
Firstly, when we inspect in detail the comparisons between
    models and observations, the clumpy models do give a significantly
improved account of the observations, and secondly the use of [\oiii]/[\oii] as 
a test for
density bounding is not valid in either type of models. The model grids, notably
for the clumpy models, overlap with, and reproduce the global 
distribution of the measurements. It is still true, as predicted from previous
types of models, that a higher [\oiii]/[\oii] ratio tends to indicate a higher 
escape
fraction of ionizing photons from a region, but the degree of ambiguity about 
this
prediction for a given example is very high, given the error bars on a given
measurement, for which the reader is referred to McCall et al. (1985).
This last conclusion is in good agreement with Iglesias--P\'aramo \& 
Mu\~noz--Tu\~n\'on (2002),
who proposed an alternative diagnostic based on the line strengths of  
[\oi]$\lambda$6300\AA\
and [\oii]$\lambda$3727\AA.

\begin{figure*}
\resizebox{0.48\hsize}{!}{\includegraphics{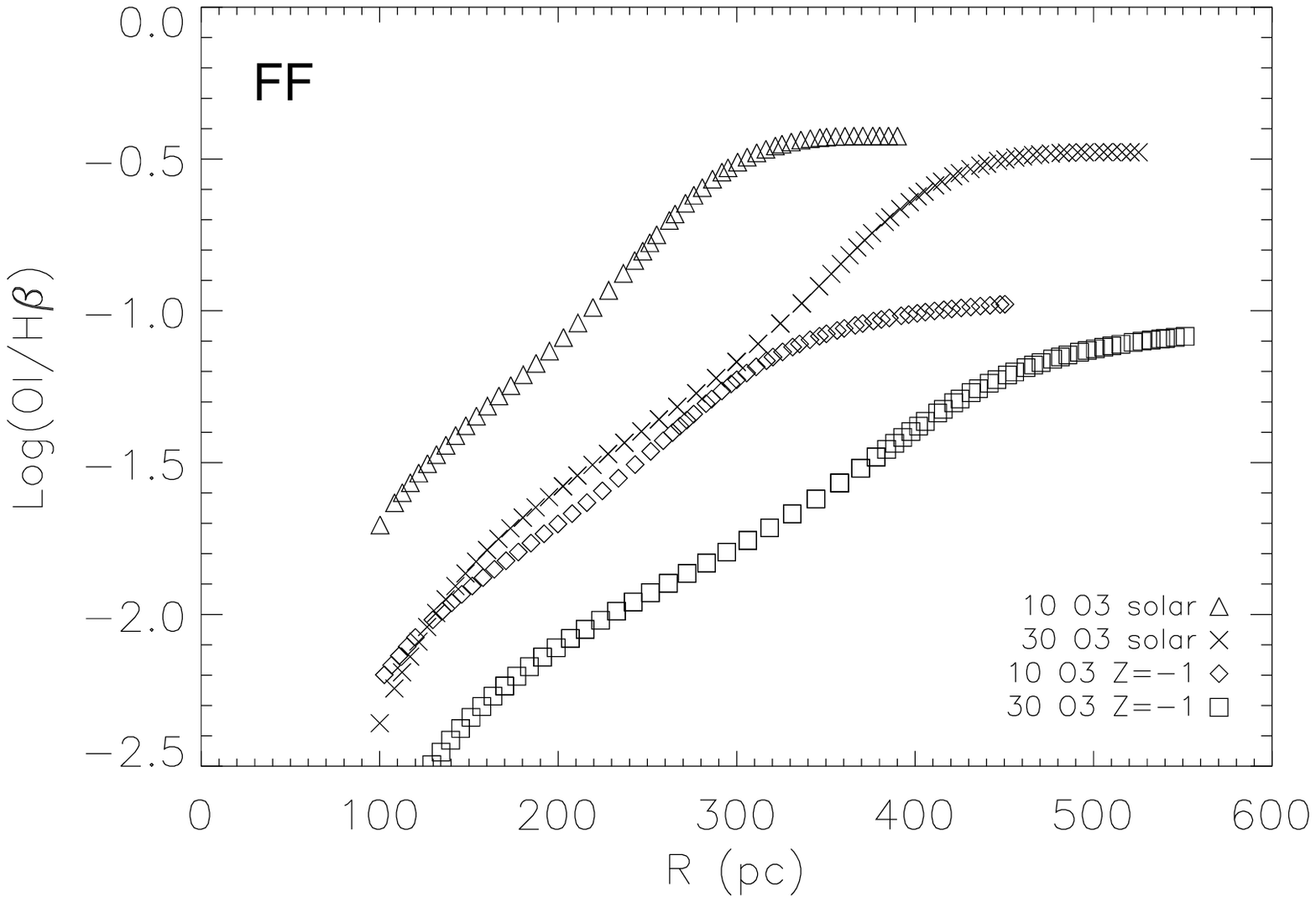}}~a.
\resizebox{0.48\hsize}{!}{\includegraphics{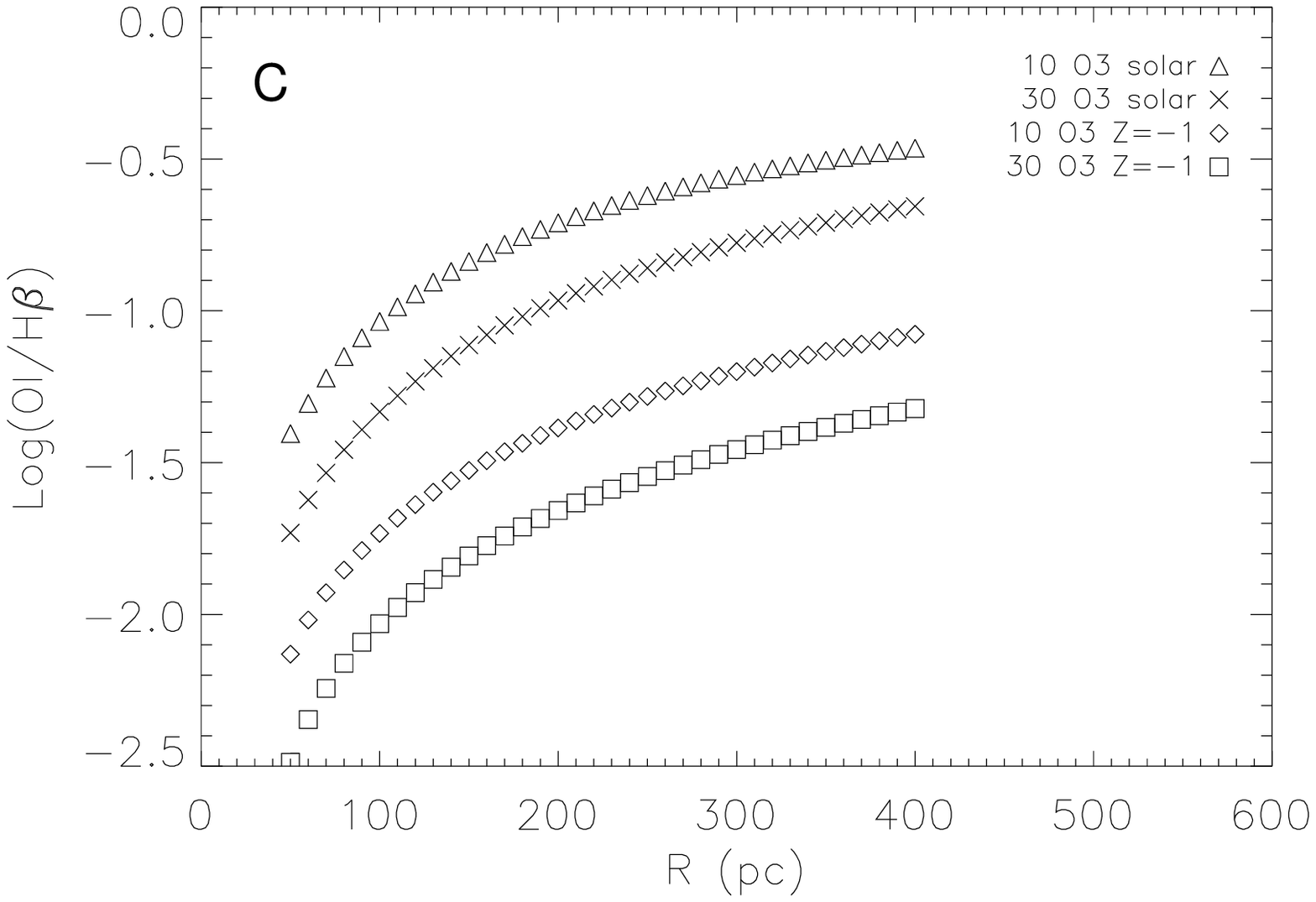}}~b.\\
\vspace{-0.4cm}
\caption{{\bf a.)} Behaviour of line ratio [\oi]6300/\hb\ versus the \hii\ 
region radius for different families of \hii\ region models in the FF 
approximation, for two different stellar contents and metallicities. As the radius 
increases, the regions change from density 
to ionization bounded. The [\oi]6300/\hb\ ratio becomes constant when the models 
become  ionization bounded.
{\bf b.)} Same as {\bf a.)} but for clumpy models. In this case the line ratio 
increases very slowly 
because the emission lines result from the sum of the contribution of all the 
clumps, which are ionization bounded.
The reason why we see a slow increase rather than a plateau is due to the fact 
that each clump has a 
different ionization parameter, depending on its distance from the ionizing stellar source.
}
\label{ox1}
\end{figure*}

In Fig.~\ref{casiana}, we compare the different predictions of clumpy (\ref{casiana}a and \ref{casiana}c) 
and FF (\ref{casiana}b and \ref{casiana}d) 
models for the [\oi] diagnostic diagram. Different values of absorption fraction 
per shell 
($f=$10$^{-2}$ and the extreme value 10$^{-3}$) and optical filling factor 
($\phi_\circ$=10$^{-3}$, and the extreme value 10$^{-4}$) have been used in the 
clumpy and 
FF models respectively.
Figs.~\ref{casiana}a and c clearly show that FF models with low escape fraction 
(dark points)
occupy the same area of the plot independently of the filling factor adopted. 
This means that
if \hii\ regions were well described by FF models, this representation would 
be valid for
density bounding diagnostics. However, clumpy 
models with high escape fraction in Fig.~\ref{casiana}b, occupy the same parameter space 
as the clumpy  models with low escape fraction in  Fig.~\ref{casiana}d. 
Therefore,
before using these diagrams, we need to make assumptions about the parameter 
$f$. 
In order to understand this behaviour, we show in Fig.~\ref{ox1}, the predicted 
[\oi]6300\AA/\hb\ line ratio as a  function of \hii\ region radius, again 
comparing 
the clumpy models (\ref{ox1}b) with ``standard''
FF models (\ref{ox1}a). 
It is a general feature of all the models that as the radius of the \hii\ region  
increases the ratio rises, but for two different reasons: in the FF case, 
there is a rapid rise  to an  asymptotic value, because as the ionizing 
photons are increasingly absorbed within the region, 
the fraction of hydrogen which remains in neutral state rises, and therefore the [\oi] emission,
   induced by a charge-exchange reaction.
However the
clumpy models tell a different story. Here there is no clear transition 
in the line intensity ratio because there is no clean transition in the 
physical properties of the regions; in other words, each clump in which emission 
lines are formed and emitted is ionization bounded itself. The distance of the 
clumps from
the ionizing sources determine different ionizing parameters, i.e. different 
values 
of the [\oi]6300\AA/\hb\ line ratio which average in a complex way giving rise 
to
the behaviour shown in Fig.~\ref{ox1}b. 
We conclude here that the distribution of data in McCall et al's plot is compatible with density 
bounding,
as density and ionization  bounded models, for both clumpy and FF, are in the same 
region of the plot. 
We have some reason to prefer clumps as for dynamical  
reasons (see next section), there does appear to be a high
    proportion of \hi\ within large \hii\ regions. 
However to 
decide definitively which is the best way to model \hii\ regions, we need to 
take into 
account other diagnostic diagrams which will be the topic of a  forthcoming paper 
(Giammanco et al. 2004).


\addtocounter{figmas}{1}

\section{Gas masses in \hii\ regions}

It is very difficult to estimate the amount of neutral hydrogen within 
an \hii\ region, because of the need to separate the contributions of 
hydrogen clouds along the line of sight to the object (and even beyond it)
from the 21~cm emission from the specific volume of hydrogen we wish to
 identify. This difficulty is serious even using Doppler techniques 
to try to isolate the emission 
from the region under study. For \hii\ regions in external galaxies the velocity 
does give us a better discriminator, but here angular resolution limits our 
ability to define the \hi\ content (and also the possible H$_2$ content) of even 
a 
large luminous \hii\ region. However there have been a few attempts to 
supplement direct 
measurements by indirect estimates of the neutral gas mass within an \hii\ 
region 
using dynamical arguments. Yang et al. (1996) estimated that in order for the 
basic 
emission line width (the width unaffected by locally expanding material, and 
which 
required careful two dimensional spectroscopy to pick out) of \ha\ from the 
region NGC~604 in M33 to be consistent with a system in virial equilibrium, the 
neutral 
gas mass had to be an order of magnitude higher than their observed mass of 
ionized gas.  
They supported their argument using an observational HI mass estimate 
by Deul \& van der Hulst (1987), and an observational estimate of the  H$_2$ mass 
by 
Viallefond et al. (1992), and they themselves showed that the total stellar mass 
is not a major fraction of the total mass of the region. A similar though less 
clear cut argument was produced by Chu \& Kennicutt (1994) for 30 
Doradus. Comparing emission line widths from many features across this huge, nearby, and 
luminous 
\hii\ region with mass estimates based on the virial theorem, they found that 
even an 
order of magnitude more \hi\ than \hii\ would not yield virial equilibrium, 
relating 
the discrepancy to dynamical energy inputs from the hot stars and supernova 
remnants. However in this case there are no corresponding literature estimates 
of 
the neutral gas (atomic or molecular ) with which to compare. It is clear 
from the above articles that if we assumed that the observational estimates 
of the ionized gas gave us the complete gas mass, i.e. that there is little or 
no neutral mass within the \hii\ regions, we would put the regions way out of 
equilibrium.
It is less clear, however that they are in fact in equilibrium 
generally, though locally perturbed by stellar energy inputs. This however, 
was the basic result of the very detailed study by Yang et al. (1996) for NGC~604.
 
\begin{figure}
\resizebox{\hsize}{!}{\includegraphics{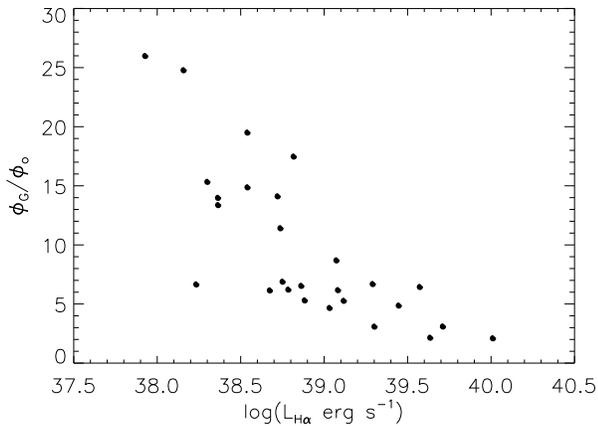}}\\[8pt]
\caption{Ratio between the geometrical and optical filling factors as a function 
of \ha\ luminosity
for \hii\ regions from the catalogue of NGC~1530 (Rela\~no et al. 2004).
$\phi_G$ has been obtained assuming spherical  clumps with radius 1~pc and a 
photon escape  fraction of 30\%. Under this hypothesis,
the mean ratio of filling factors is $\sim10$, while assuming a  photon escape  
fraction of 50\% the mean ratio is $\sim6$.
}
\label{mass}
\end{figure}

In this context we see that we could use our clumpy models to estimate 
neutral gas masses in \hii\ regions purely on the basis of ionization 
equilibrium, once the basic postulates about the clump properties had been laid 
down. The calculation is straightforward. Our model allows us to estimate the 
fraction of a clump which is ionized at any chosen distance from the ionizing 
source, so once we know, or assume, the distribution of clumps within the region 
as a 
whole we can compute the neutral gas fraction. The results of calculations of 
this 
type for a set of observed \hii\ regions in the barred spiral NGC~1530 are shown 
in Fig.~\ref{mass}.
The results shown are based on the specific assumption that a constant 
fraction of 30\% of the ionizing radiation from the stellar sources escapes from 
all the \hii\ regions, but the qualitative result would not change if we used a 
different basis for the computation (for example increasing this escape factor 
to 50\% in 
fact decreases the neutral/ionized gas mass ratio less then factor 2). The 
graph  shows the ratio of true clump filling factor, $\phi_G$, which is the 
fractional volume of an \hii\ region occupied by dense gas clumps, to the
optical filling factor $\phi_\circ$, which is the fraction occupied by the 
ionized portions of the gas clumps.
We can see from Fig.~\ref{mass} that ratios of an order of magnitude  between 
the neutral gas and ionized gas masses are indicated in these models, as a 
natural consequence of the clumpy nature of the medium. This gives a firmer 
basis for the use of this factor in making the calculation of the gas mass in 
NGC~604, and implies that a virial equilibrium solution for the line widths 
observed by Yang et al. makes good physical sense. In a forthcoming paper 
(Rela\~no 
et al. 2004, A\&A, submitted) we will present a general study of the topic of 
the 
internal kinematics of \hii\ regions based on observations of complete 
populations 
of regions in the discs of spiral galaxies, in which one of our conclusions is 
that the widths, $\sigma$ of the integrated \ha\ emission from the majority of 
the \hii\  regions observed in external galaxies are broader than those expected 
for systems in 
virial equilibrium, but that the lower limiting envelope in sigma for the 
whole population, plotted in a diagram of $\sigma$ v. \ha\ luminosity, 
represents a locus 
of regions which are virialized.Were it not for the major contribution of the 
neutral gas in the clumps the regions would all appear to be gravitationally 
unbound.    

\addtocounter{figmas}{1}

 We can thus conclude, at least tentatively, that measurements of the
internal dynamics of \hii\ regions, supplemented by direct estimates
of \hi\ in the few cases where this is possible, give good support to
the use of our clumpy models to interpret the emission parameters.
This support is quantifiable in terms of the fractional estimates
of neutral gas mass obtained dynamically compared with the
fractional estimates based on clumpy models with reasonable radiative
parameters. Lastly it is relevant to point out that Rela\~no et al.
(2004) have shown that the stellar mass cannot make up more than a
few percent of the total \hii\ region mass for these luminous regions,
and so does not materially affect the dynamical balance of a
region. 

\section{Conclusions}

   The aim of this paper has been to offer an initial alternative version
to the conventional inhomogeneous models for \hii\ regions.
The classical filling factor approximation assumes that the denser clumps are optically 
thin, 
and this gives rise to a statistically modified Str\"omgren sphere with a well 
defined
boundary in ionization bounded conditions. The use of an optically thick
approximation gives some substantially different predictions. We have shown
in particular, the differences implied in the diagnostic diagram of
[\oiii]/[\oii] v. ([\oiii]+[\oii])/\hb, a diagram used, inter alia, for 
separating regions which are ionization bounded from those which are density 
bounded.
The optically thick clumpy model gives predictions in rather better agreement
with the observed data than the traditional filling factor models, but more
critically, we show that in neither model is the separation of the two
conditions (ionization  and density bounding) very clean, and in
the ``clumpy" model there is complete overlap, rendering the diagram
ineffective as a diagnostic. We apply the same exercise to the more
recently suggested [\oi]/\hb\  v. [\oii]/\hb\ diagram, showing that in  this
case the diagnostic would function adequately for the traditional
models, but for the optically thick clumpy models the results would be ambiguous.
The models described here, though still geometrically quite schematic,
do appear to give an improved account of the radial profiles
of \hii\ regions in surface brightness. They also predict as a natural
consequence that \hii\ regions should contain a majority fraction of \hi,
and this prediction seems to be borne out by observations of the internal
dynamics of \hii\ regions.

\begin{acknowledgements}
Thanks are owed to Tom Abel for scientific discussions on radiative transfer in
diffuse media. We also thank the referee, Manuel Peimbert for helpfull suggestions
which helped us to improve the article.
This work was supported by the Spanish DGES (Direcci\'on General de
Ense\~nanza Superior) via Grants PB91-0525, PB94-1107 and PB97-0219
and by the Ministry of Science and Technology via grant AYA2001-0435. 
A.~Z. acknowledges support from 
the Consejer\'\i a de Educaci\'on y Ciencia de la Junta de Andaluc\'\i a, Spain.
A.~Z. and C.~G. acknowledge support from the Isaac Newton Group (ING) during the preparation of
this article. The JKT is operated on the island of La Palma by the ING in the
Spanish Observatorio del Roque de los Muchachos of the Instituto de Astrof\'\i sica de
Canarias. 
\end{acknowledgements}

\end{document}